\documentclass[aps,preprint,floatfix,color]{revtex4}
\usepackage{graphicx}
\usepackage{color}
\usepackage{epstopdf}
\usepackage{bm}
\usepackage{amssymb}

\newcommand{\bea}{\begin{eqnarray}}
\newcommand{\eea}{\end{eqnarray}}
\newcommand{\be}{\begin{equation}}
\newcommand{\ee}{\end{equation}}

\begin{document}
\title{Thin Films of Chiral Motors}
\author{M. Strempel$^{1,2}$, S. F\"urthauer$^{1,2}$, S. W. Grill$^{1,2}$, F. J\"ulicher$^{1}$}
\affiliation{$^{1}$Max Planck Institute for the Physics of Complex Systems, N\"othnitzer Stra\ss e 38, 01187 Dresden, Germany}
\affiliation{$^{2}$Max Planck Institute of Molecular Cell Biology and Genetics, Pfotenhauerstr. 108, 01307 Dresden, Germany}

\begin{abstract}
Hydrodynamic flows in biological systems  are often generated by active chiral processes near or on surfaces. Important examples are beating cilia, force generation in actomyosin networks, and motile bacteria interacting with surfaces. Here we develop a coarse grained description of active chiral films that captures generic features of flow and rotation patterns driven by chiral motors. We discuss force and torque balances within the film and on the surface and highlight the role of the intrinsic rotation field. We arrive at a two dimensional effective theory and discuss our results in the context of ciliary carpets and thin films of bacterial suspensions.
\end{abstract}

\maketitle

The building blocks of biological systems are chiral molecules such as proteins and DNA.
The emergent collective behaviors of these molecular components give rise to 
active dynamic processes in cells and tissues which can also reflect these molecular chiralities.
A key example is the breaking of left-right symmetry in organisms with a well-defined handedness during development
\cite{wolpert98,vand09,henl08} (i.e. the heart is on the left side in humans).
Many pattern forming events in biological systems are governed by active chiral processes.
For example, it has been shown that the rotating beat of cilia,
which drives chiral hydrodynamic flows, is at the basis of left-right symmetry breaking in vertebrate animals  \cite{nona98,nona05,buce05}. 
In addition, active chiral processes have been observed in the cytoskeleton of cells \cite{dani06,xu07}. The cytoskeleton is a gel-like network of elastic chiral filaments and other components such as motor proteins.
The interactions of motors and filaments drive movements and intracellular flows \cite{mayer10,meen08}, which can have chiral asymmetries \cite{sase97,dani06,hilf07, vilf09}. 

\begin{figure}[t]
\includegraphics[width=.55\textwidth]{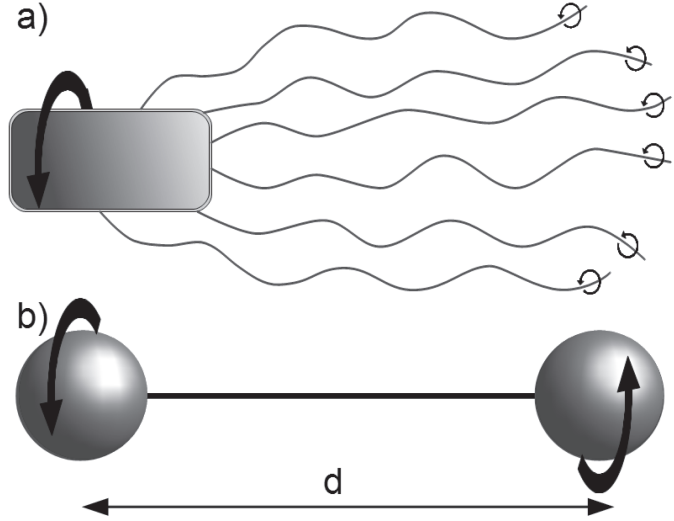}
\caption{ (a) Schematic representation of a swimming bacterium with rotating flagella. (b) Chiral  
motor consisting of two counterrotating spheres at distance $d$. Arrows indicate senses of rotation.}
\label{fig:nematic_motor}
\end{figure}

Active chiral processes are typically observed on surfaces or at interfaces. Key examples are
carpets of beating cilia  driving hydrodynamic flows parallel to the surface, on which they are
attached  \cite{gibb81, bray00}, and rotating motors on a surface \cite{noji97, lenz03,uchi10}. Many different beating patterns of cilia exist, which in general are chiral and often exhibit
rotating movements, as is the case for the cilia that govern the left-right symmetry breaking of organisms \cite{buce05, smit08}.
 Many microorganisms  possess carpets of cilia on their outer surface, which
 are used for self-propulsion along helical trajectories in a fluid \cite{bray00}.
Recently, the collective behavior of swimming \it E. coli \rm bacteria
on solid surfaces was studied \cite{berg11}. These bacteria possess rotary motors, which drive
the rotation of helical flagella relative to the cell body. This relative rotation provides a  motor
for self-propulsion in a fluid.
Due to conservation of angular momentum, cell body and flagella exert equal but 
opposite torques on the surrounding fluid, see Fig.~{\ref{fig:nematic_motor}(a)}. 
Thus, the bacteria act as torque dipoles.
Such flagellated \textit{E.~coli}  were  placed
on a solid agar surface~\cite{berg11}, where they produced a thin fluid layer that promotes their motility.
Some bacteria attach to the surface while others swim in the film.
Those bacteria which were attached to the surface also exerted torques on the solid substrate.
Large scale chiral flow patterns were reported as a result of bacterial activity~\cite{berg11}. 
The bacterial suspension therefore represents an internally driven active chiral film supported by a surface.

Active fluids and gels have been studied in the framework of hydrodynamic 
theories both as a paradigm for the
cell cytoskeleton and for suspensions of active swimmers~\cite{simh02,live03,aran03,hatw04,krus04,krus05}. 
Such approaches are based on liquid crystal hydrodynamics~\cite{eric60,lesl66,mart72,dege95,stark03,stark05} driven out equilibrium by internal active processes. It has been shown that stresses generated by active processes can give rise to a rich variety of
dynamic patterns and flows \cite{voit05,salb09,mayer10,bois11,giom11,elge11}.
Similar approaches have also been used for the study of granular systems \cite{bocquet01,aran06}. 
Chiral effects have so far been discussed mainly in passive systems and in  
granular gases \cite{tsai05}.
A study of active chiral processes in fluids and gels is so far lacking.

Here, we develop a generic theory for active chiral films. We consider a thin film of a suspension of chiral motors on a solid surface.
 Our work is based on a general discussion of the bulk properties of active
chiral fluids~\cite{fuer11}. 
Active chiral processes are introduced as torque dipoles arising from 
counterrotating objects, such as 
the cell body and the flagella of bacteria, see Fig.~{\ref{fig:nematic_motor}(a)}.

We start by discussing the bulk properties of a fluid with mass density $\rho$ and
center of mass velocity $\bf{v}$ in which active  
processes take place. Linear momentum and angular momentum conservation
can be expressed as
\begin{eqnarray}\label{eq:force_balance}
\partial_t (\rho v_\alpha)&=&\partial_\beta\sigma^{tot}_{\alpha\beta} +f^{ext}_{\alpha} \quad, \\
\partial_t l^{tot}_{\alpha\beta}&=&\partial_\gamma M^{tot}_{\alpha\beta\gamma} +\tau^{ext}_{\alpha\beta}+ r_\alpha f^{ext}_\beta -r_\beta f^{ext}_\alpha\quad,
\end{eqnarray}
where Einstein's summation convention is implied.
Here $\rho v_\alpha$ is the momentum density. The density of 
angular momentum $l^{tot}_{\alpha\beta}$
is described by an antisymmetric second rank tensor. Externally 
applied force and torque densities are denoted
by $f^{ext}_\alpha$ and $\tau^{ext}_{\alpha\beta}$, respectively, 
$r_\alpha$ is a position vector, $\sigma^{tot}_{\alpha\beta}$ is the 
total stress and $M^{tot}_{\alpha\beta\gamma}$ is the total angular momentum flux. 
 The total stress $\sigma^{tot}_{\alpha\beta} = -P\delta_{\alpha\beta}
+ \sigma^{s}_{\alpha\beta}+\sigma^{a}_{\alpha\beta}$ describes 
momentum fluxes and can be decomposed in the isotropic pressure $P$,
the symmetric traceless stress $ \sigma^{s}_{\alpha\beta}$ and the antisymmetric stress $\sigma^{a}_{\alpha\beta}$.  
 The total angular momentum density consists of an orbital part 
 $\rho (r_\alpha v_\beta-r_\beta v_\alpha)$ and
a spin angular momentum density 
$l_{\alpha\beta}=l^{tot}_{\alpha\beta}-\rho (r_\alpha v_\beta-r_\beta v_\alpha)$, see \cite{eric60,lesl66,tsai05,stark05}.
Similarly, the  total angular momentum flux
$M^{tot}_{\alpha\beta\gamma}=
M_{\alpha\beta\gamma}+(r_\alpha\sigma_{\beta\gamma}^{tot}-r_\beta\sigma_{\alpha\gamma}^{tot})$
is the sum of  fluxes $M_{\alpha\beta\gamma}$ and $r_\alpha\sigma^{tot}_{\beta\gamma}-r_\beta\sigma^{tot}_{\alpha\gamma}$ of spin and orbital angular momentum, respectively. Note that the spin angular momentum density
$l_{\alpha\beta}$ and the spin angular momentum flux $M_{\alpha\beta\gamma}$ do not depend
on the choice of coordinate system.
We define an effective rate of intrinsic rotation 
$\Omega_{\alpha\beta}$ of local volume elements via $l_{\alpha\beta}=I_{\alpha\beta\gamma\delta}\Omega_{\gamma\delta}$, 
where $I_{\alpha\beta\gamma\delta}$ is the moment of inertia tensor per unit volume \cite{stark05}.
The  balance of the local torque then reads
\begin{eqnarray}\label{eq:torque_balance}
\partial_t l_{\alpha\beta}&=& - 2\sigma^a_{\alpha\beta} +\partial_\gamma M_{\alpha\beta\gamma} +\tau^{ext}_{\alpha\beta}\quad .
\end{eqnarray}  
Eq.~{(\ref{eq:torque_balance})} shows that spin angular momentum is not conserved but can be exchanged with orbital angular momentum 
by antisymmetric stress. 

Active chiral processes introduce contributions to the total stress and to fluxes of
spin angular momentum. We discuss the effect of an active chiral process by considering
a torque dipole $\tau_{\alpha\beta}(\mathbf{r})$ in a fluid, located at $\mathbf{r}=0$, see Fig.~{\ref{fig:nematic_motor}(b)}.
 This torque dipole is built from two torque monopoles $\pm q \epsilon_{\alpha\beta\gamma}p_\gamma $, of strength $q$
separated by a small distance $d$ in the direction of the torque axis given by the unit vector $\bf p$:
\begin{eqnarray}\label{eq:torque_dipole}
\tau_{\alpha\beta}&=&q\epsilon_{\alpha\beta\gamma}{p_\gamma}\left(\delta(\mathbf{r}
-\frac{{d}}{2}{\bf p})-\delta(\textbf{r}
+\frac{{d}}{2}{\bf p})\right)\nonumber\\
&\simeq& -qd\epsilon_{\alpha\beta\nu}{p_\nu}p_\gamma\partial_\gamma\delta({\bf r})\quad .
\end{eqnarray}
Note that $\tau_{\alpha\beta}$ is invariant under the transformation 
$\bf{p}\rightarrow-\bf{p}$, which implies a nematic character. 

Active internal torques and forces are introduced similarly to external ones. Therefore the torque dipole $\tau_{\alpha\beta}$ can be interpreted as an active contribution  
$M^{act}_{\alpha\beta\gamma}=-qd\epsilon_{\alpha\beta\delta}{p_\delta}p_\gamma \delta({\bf r})$
to the spin angular momentum flux, which obeys $\partial_\gamma M_{\alpha\beta\gamma}^{act}=\tau_{\alpha\beta}$.
In a suspension of many identical torque 
dipoles at positions ${\bf r}^{(i)}$ and with orientations ${\bf p}^{(i)}$ the active angular momentum fluxes are
\begin{eqnarray}
M^{act}_{\alpha\beta\gamma}&=&- qd\sum_i  \epsilon_{\alpha\beta\delta}p^{(i)}_\delta p^{(i)}_\gamma\delta({\bf r}-{\bf r}^{(i)}) \nonumber\\\label{eq:active_angmomflux}
 &\simeq& \zeta \epsilon_{\alpha\beta\delta}p_\delta p_\gamma+\zeta^\prime\epsilon_{\alpha\beta\gamma}\quad,
\end{eqnarray}
where $\zeta=-Sqd n(\bf{r})$ and $\zeta^\prime =(S-1)qd n(\textbf{r})/3$ describe the strength of a nematic and an isotropic active contribution to $M_{\alpha\beta\gamma}$, respectively, and $n(\mathbf{r})$ is the dipole density. 
Here and below $\mathbf{p}$, with ${\bf p}^2=1$, denotes a coarse grained nematic director and $S$ is a nematic order parameter which obeys $S(p_\alpha p_\beta-(1/3)\delta_{\alpha\beta})=<p^{(i)}_\alpha p^{(i)}_\beta>-(1/3)\delta_{\alpha\beta}$,
where the average is taken over a  volume element \cite{dege95}.

The constitutive equation for the spin angular momentum flux then has the form
\begin{eqnarray}\label{eq:ang_mom_flux}
M_{\alpha\beta\gamma}=\kappa\partial_\gamma\Omega_{\alpha\beta}+\zeta \epsilon_{\alpha\beta\delta}p_\delta p_\gamma+\zeta^\prime\epsilon_{\alpha\beta\gamma}\quad .
\end{eqnarray} 
Here the term proportional to the phenomenological coefficient $\kappa$ describes passive dissipative processes. Furthermore we ignore other passive couplings \cite{stark05,fuer11}.  Here and below we have neglected inertial terms. We complement Eq.~{(\ref{eq:ang_mom_flux})} by the constitutive equations for the stresses of a passive incompressible fluid: 
\begin{eqnarray}\label{eq:sym_stress}
 \sigma_{\alpha\beta}^s&=&2\eta \tilde u_{\alpha\beta}\quad,\\\label{eq:antisym_stress}
\sigma^a_{\alpha\beta}&=&2\eta^\prime(\Omega_{\alpha\beta}-\omega_{\alpha\beta})\quad ,
\end{eqnarray}
where $\tilde u_{\alpha\beta}$ is the traceless part of the strain rate $u_{\alpha\beta}=\left(\partial_\alpha v_\beta+\partial_\beta v_\alpha\right)/2$ and  $\omega_{\alpha\beta}=\left(\partial_\alpha v_\beta-\partial_\beta v_\alpha\right)/2$. Here, $\eta$ is the shear viscosity and $\eta^\prime$ is a rotational viscosity. The pressure $P$ plays the role of a Lagrange multiplier and is imposed by the incompressibility condition $\partial_\gamma v_\gamma=0$. Other coupling terms, known to exist in liquid crystals, have been neglected for simplicity.  

\begin{figure}[t]
\includegraphics[width=.7\textwidth]{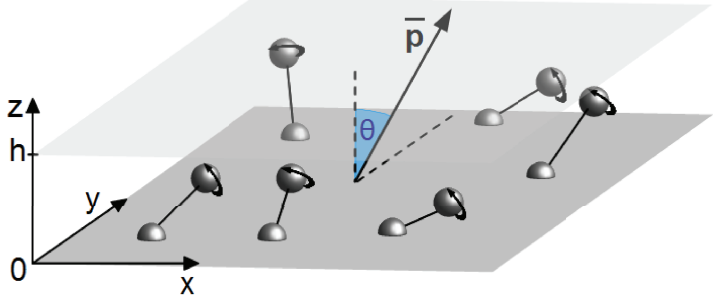}
\caption{Schematics of a thin fluid film of height $h$ that contains chiral motors. 
Motors are torque dipoles that consist of counterrotating spheres (see Fig.~{\ref{fig:nematic_motor}(b)}), one of which is attached to the surface. 
The other rotates as indicated by the arrows. The average motor direction is described by the
vector $\bar\mathbf{p}$ which is tilted at an angle $\theta$ in $y$-direction.}
\label{fig:motors_on_surface}
\end{figure}

We now study the stresses and flows generated by active chiral processes in a thin fluid film  of height $h$ on a solid substrate, see Fig.~{\ref{fig:motors_on_surface}}. Note that the film is not symmetric with respect to $z\to h-z$ because the two surfaces at $z=0$ and $z=h$ differ.  
Integrating the force balance  Eq.~{(\ref{eq:force_balance})} we obtain
\begin{equation}\label{eq:2d_force_balance}
0=\frac{1}{h}\int\limits_0^h \mathrm{d}z \partial_\beta \sigma^{tot}_{i\beta}= 
\partial_j \bar\sigma^{tot}_{i j}+\frac{1}{h}\sigma^{tot}_{i z}|_0^h \quad,
\end{equation}
where the bar denotes an average over the film height, 
$\bar\sigma^{tot}_{i\beta}\equiv(1/h)\int_0^h \mathrm{d}z\sigma_{i\beta}^{tot}$, and
$\sigma^{tot}_{i z}|_0^h\equiv\sigma^{tot}_{i z}(z=h)-\sigma^{tot}_{i z}(z=0)$, where $i=x,y$ and $j=x,y$, see Fig.~{\ref{fig:motors_on_surface}}.
The shear stress at the film surfaces is written as
\begin{eqnarray}
\frac{1}{h}\sigma^{tot}_{i z}|_0^h =-\xi_t \bar v_i +\xi_\Omega \bar\Omega_{iz}\quad ,
\end{eqnarray}
which describes friction forces due to relative motion with the substrate by the coefficient $\xi_t$ and due to relative rotation with the substrate by the coefficient $\xi_\Omega$.
From the torque balance Eq.~{(\ref{eq:torque_balance})} and ignoring inertial terms, we find
\begin{eqnarray}
0&=&-2\bar\sigma^a_{\alpha\beta} +\partial_j \bar M_{\alpha\beta j} + \frac{1}{h} M_{\alpha\beta z}|_0^h\quad.
\end{eqnarray}
The torque densities on the surfaces are
\begin{eqnarray}
\frac{1}{h} M_{\alpha\beta z}|_0^h &=& -\xi_r \bar\Omega_{\alpha\beta} +\zeta_s\epsilon_{\alpha\beta \delta}\bar p_z \bar p_\delta +\zeta^\prime_s\epsilon_{\alpha\beta z}\quad ,
\end{eqnarray}
where frictional torques are described by the coefficient $\xi_r$. The torque exerted by active chiral processes on the boundaries is described by the coefficients
$\zeta_s$ and $\zeta^\prime_s$.
 Averaging over the height of the film also leads to new coefficients for the active in-film angular momentum fluxes $\bar M_{\alpha\beta j}=\zeta_f\epsilon_{\alpha\beta\delta} \bar p_\delta \bar p_j+\zeta^\prime_f\epsilon_{\alpha\beta j} $.
 Finally, the externally applied shear stress on the boundaries is $\sigma_{iz}^{ext}\approx\bar\sigma_{iz}=(\eta+\eta^\prime)v_i|_0^h/h+2\eta^\prime\bar\Omega_{iz}$. In the absence of external shear stresses $v_i|_0^h=-2h\eta^\prime\bar\Omega_{iz}/(\eta+\eta^\prime)$. Similarly, $\bar\sigma_{zz}=P^{ext}$ is the externally applied pressure and therefore $\bar P= -2\eta\partial_i \bar v_i$. 
We obtain equations for the flow and rotation field, 
\begin{eqnarray}\label{eq:2d_force_balance_eqofmotion}
0&=&\partial_j \left[ (3\eta-\eta^\prime) \partial_i \bar v_j +(\eta+\eta^\prime) 
\partial_j \bar v_i \right]+ 2\eta^\prime\partial_j \bar\Omega_{ij}-\xi_t \bar v_i+\xi_\Omega\bar\Omega_{i z} \quad, \\
\label{eq:2d_torque_balance_eqofmotionxy}
0&=&-(4\eta^\prime+\xi_r)\bar\Omega_{xy}+2\eta^\prime (\partial_x \bar v_y-\partial_y \bar v_x)+\kappa\partial^2_j\bar\Omega_{xy}+\partial_j\zeta_f\epsilon_{xyz}\bar p_z \bar p_j +\zeta_s\epsilon_{xyz}\bar p_z^2+\zeta^{\prime}_s\epsilon_{xyz}\quad ,\\
\label{eq:2d_torque_balance_eqofmotioniz}
0&=&-\left(\frac{4\eta\eta^\prime}{\eta+\eta^\prime}+\xi_r\right)\bar\Omega_{iz}+\kappa\partial^2_j\bar\Omega_{iz}+\partial_j\zeta_f\epsilon_{iz\delta}\bar p_\delta \bar p_j+\partial_j\zeta^{\prime}_f\epsilon_{iz j} +\zeta_s\epsilon_{iz\delta}\bar p_\delta \bar p_z\quad ,
\end{eqnarray}
where we used the thin film approximation $\partial_i v_z \ll \partial_z v_i$.

Motivated by rotating cilia or bacteria attached to a surface, we now consider the case of a thin active film containing chiral motors that  are on average aligned
along the vector $\bar\mathbf{p}=(\cos\phi\sin\theta,\sin\phi\sin\theta,\cos\theta)$. 
Here we have introduced the angle $\theta$ which describes the average tilt with respect to the surface normal vector and the angle $\phi$ which specifies the tilt direction with respect to the $x$-axis. 
We first consider a homogeneous distribution of motors in an infinite system where all spatial derivatives vanish.
In this case Eq.~{(\ref{eq:2d_force_balance_eqofmotion})} reads
$\bar v_i=(\xi_\Omega/\xi_t) \bar\Omega_{iz}$.
Using Eq.~{(\ref{eq:2d_torque_balance_eqofmotioniz})} we find 
\begin{eqnarray}\label{eq:hom_carpet_velocity}
\bar v_i&=&\frac{\xi_\Omega\zeta_s}{\xi_t}\ \frac{\epsilon_{izj}\bar p_j\bar p_z}{\xi_r+4\eta\eta^\prime/(\eta+\eta^\prime)}\quad .
\end{eqnarray}
A flow with a velocity $ | \bar{\mathbf{v}} | \propto \sin(2 \theta )$
is generated. The flow direction is
perpendicular to the direction of tilt with respect to the surface normal vector. This case describes the collective generation of flow by carpets of cilia on a surface \cite{buce05,nona05,smit08}. 
Note that $\xi_\Omega$ and therefore also the generated flow vanish in films that are symmetric with respect to  $z\to h-z$. Ciliary carpets do have this asymmetry and can therefore generate flows \cite{buce05}. 
Furthermore, we find from Eq.~{(\ref{eq:2d_torque_balance_eqofmotionxy})} an intrinsic rotation rate $\bar\Omega_{xy}=(\zeta_s \cos^2\theta+\zeta^{\prime}_s)/(4\eta^\prime+\xi_r)$ generated by the rotating motors with vorticity $\bar\omega_{xy}=0$ in the hydrodynamic flow field. 

\begin{figure}[h]
\includegraphics[width=.65\textwidth]{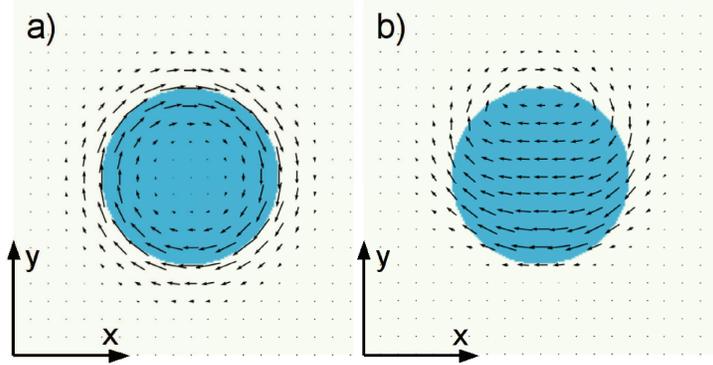}
\caption{Flow fields (vectors) generated by a circular patch of chiral motors (blue).  
a) Average motor axis $\bar\mathbf{p}$ is perpendicular to the surface, i.e. $\theta=0$. 
b) Average motor axis $\bar\mathbf{p}$ tilted at angle $\theta=\pi/4$ in $y$-direction. 
Flow fields were determined numerically in a square box of size $L$ with periodic boundary conditions.
Parameter values are
$\eta^\prime=\eta=h^2\kappa=h^2\xi_t=-h\xi_\Omega=\xi_r$, $\bar\zeta=-h\zeta_s =10\bar\zeta^\prime=-10h\zeta^\prime_s$, $L=20h$.}
\label{fig:flowfields}
\end{figure}

As a second example we consider a circular patch of radius $R$ that contains active motors. This is described by position dependent coefficients $\zeta_{s}(\mathbf{r})=\zeta_{s}\Theta(R-|\mathbf{r}|)$ and by similar expressions for $\zeta^{\prime}_s$, $\zeta_f$ and $\zeta^{\prime}_f$, where $\Theta(r)$ is the Heaviside function. 
We numerically solve Eqs.~{(\ref{eq:2d_force_balance_eqofmotion})}, (\ref{eq:2d_torque_balance_eqofmotionxy}) 
and (\ref{eq:2d_torque_balance_eqofmotioniz}) with periodic boundary conditions for boxsize $L=4R$ using Fourier transforms,  see Fig.~{\ref{fig:flowfields}}.
The velocity field for a tilt angle $\theta=0$ is displayed in Fig.~{\ref{fig:flowfields}(a)}.
The patch generates a chiral flow field driven by the intrinsic rotations $\bar\Omega_{xy}$. The stream lines are concentric circles. The flow velocity is maximal at the edge of the patch and decays exponentially outside. There is no net transport across the patch.
If the motors are tilted with $\theta\ne0$ along the $y$ axis, a net transport in $x$ direction across the patch with velocity proportional to  $\sin(2\theta)$ appears in addition to the circular flow, see Fig.~{\ref{fig:flowfields}(b)}. 

We have developed 
a coarse grained description of flow patterns generated in thin active films in which chiral processes take place.
We have shown that if the chiral motors are tilted with respect to the surface normal vector, directed flows can be generated over large
distances in a direction perpendicular to the tilt direction. The velocity of this net flow is maximal for tilt angles
of $\theta=45^\circ$. This result accounts for the generation of net flows and left-right symmetry breaking 
by carpets of cilia in the mouse ventral node \cite{nona98,nona05,buce05} and Kupffer's vesicle in zebrafish \cite{essn05}. 
Note that in both cases a tilt angle with respect to the surface normal can be defined
\cite{smit08}. Interestingly, reported tilt angles vary between $30^\circ$ and $50^\circ$ \cite{okada05}, which is close to the angle of maximum transport velocity. 
Moreover in the case of finite patches of chiral motors, intrinsic rotation rates drive chiral flows along the
edge of the patch. This phenomenon was reported in recent experiments on bacterial films on solid surfaces \cite{berg11}.

Our theory highlights the role of the intrinsic rotation field $\Omega_{\alpha\beta}$ for active chiral processes. 
In particular we find  net flows without vorticity generated
by intrinsic rotations in a homogeneous system, see Eq.~{(\ref{eq:hom_carpet_velocity})}.  Note that  in passive bulk fluids $\Omega_{\alpha\beta}$ converges  
to the flow vorticity after a relaxation time
that in general is short \cite{stark05}.  In films of chiral motors, however, the intrinsic rotation rate differs from the vorticity even at steady state, and can create effects near interfaces and surfaces. Our work shows that the hydrodynamic flow velocity $\mathbf{v}$ and the intrinsic rotation rate $\Omega_{\alpha\beta}$ in active chiral films are coupled by the active processes and by the boundary conditions.
This can give rise to complex flow patterns generated by carpets of chiral motors.

 Our theory permits a simplified description of the flows generated by ciliary carpets. 
It will be interesting
to expand this theory to include phase relationships between adjacent motors to capture the self-organization
of beating cilia in metachronal waves \cite{bray00}.

\end{document}